\documentclass[sigconf]{acmart}
\AtBeginDocument{%
  \providecommand\BibTeX{{%
    \normalfont B\kern-0.5em{\scshape i\kern-0.25em b}\kern-0.8em\TeX}}}

\setcopyright{acmcopyright}
\copyrightyear{2024}
\acmYear{2024}
\setcopyright{acmlicensed}\acmConference[CHIIR '24]{Proceedings of the 2024 ACM SIGIR Conference on Human Information Interaction and Retrieval}{March 10--14, 2024}{Sheffield, United Kingdom}
\acmBooktitle{Proceedings of the 2024 ACM SIGIR Conference on Human Information Interaction and Retrieval (CHIIR '24), March 10--14, 2024, Sheffield, United Kingdom}
\acmPrice{15.00}
\acmDOI{10.1145/3627508.3638305}
\acmISBN{979-8-4007-0434-5/24/03}

\acmConference[ACM SIGIR]{CHIIR}{March 10-14, 2024}{Sheffield, UK}
\acmSubmissionID{}
\usepackage{graphicx}
\usepackage{caption}
\usepackage{subcaption}
\usepackage{lipsum}
\usepackage{enumitem}
\usepackage{float}

\begin{document}

\title{Comparing Traditional and LLM-based Search for Image Geolocation}

\author{Albatool Wazzan}
\orcid{0000-0002-0181-700X}
\affiliation{%
  \department{Dept of Computer \& Info Sciences}
  \institution{Temple University}
  \streetaddress{1925 North 12th Street}
  \city{Philadelphia}
  \country{USA}
  \postcode{19122}
}
\email{albatool.wazzan@temple.edu}

\author{Stephen MacNeil}
\orcid{0000-0003-2781-6619}
\affiliation{%
  \department{Dept of Computer \& Info Sciences}
  \institution{Temple University}
  \streetaddress{1925 North 12th Street}
  \city{Philadelphia}
  \country{USA}
  \postcode{19122}
}
\email{stephen.macneil@temple.edu}

\author{Richard Souvenir}
\orcid{0000-0002-6066-0946}
\affiliation{%
   \department{Dept of Computer \& Info Sciences}
  \institution{Temple University}
  \streetaddress{1925 North 12th Street}
  \city{Philadelphia}
  \country{USA}
  \postcode{19122}
}
\email{souvenir@temple.edu}

\begin{abstract}
Web search engines have long served as indispensable tools for information 
retrieval; user behavior and query formulation strategies have been well studied.
The introduction of search engines powered by large language models (LLMs) 
suggested more conversational search and new types of query strategies.
In this paper, we compare traditional and 
LLM-based search for the task of image geolocation, i.e., 
determining the location where an image was captured. Our work examines user interactions, with a particular focus on query formulation strategies. In our study, 60 participants were assigned either traditional or LLM-based search engines as assistants for geolocation. Participants using traditional search more accurately predicted the location of the image compared to those using the LLM-based search. Distinct strategies emerged between users depending on the type of assistant. Participants using the LLM-based search issued longer, more natural language queries, but had shorter search sessions. When reformulating their search queries, traditional search participants tended to add more terms to their initial queries, whereas participants using the LLM-based search consistently rephrased their initial queries.

\end{abstract}

\maketitle




\section{Introduction}

For decades, web search engines have served as the de facto reference tool for a wide range of tasks.
In fact, it has been demonstrated that humans have been trained to optimize keyword-based searching using query formulations not typically used in natural language~\cite{lewandowski2008search}.
Advancements in artificial intelligence (AI) have driven the 
emergence of large language models (LLM), such as 
BERT~\cite{devlin2019bert}, GPT-3~\cite{brown2020language}, and their successors. These models have served as the foundation for numerous applications, ranging from text generation, translation, to question answering, multi-step reasoning, and complex problem solving~\cite{wei2022emergent}. Recently, these tools have been combined with web search to enable a new mode of LLM-powered conversational search. Unlike keyword-based search, this integration allows users to engage in a natural, interactive conversation with the LLM-powered search engine, as if they were interacting with a knowledgeable assistant. This conversational mode has the potential to improve user experiences across various domains. While previous work has explored how people engage in sense-making and constructing mental models of traditional search engines~\cite{thomas2019investigating}, the adaptation of these models to LLM-based search remains open to inquiry. 

To compare traditional and LLM-based search, we consider the task of image geolocation -- identifying the location in which an image was captured, an important task with applications in forensics, law enforcement, and journalism. This task has historically been performed by expert image analysts, using increasingly sophisticated reference tools as they became available. Fully automated computer vision approaches~\cite{hays2008im2gps, Weyand_2016,7299135} have been developed; these approaches typically rely on the visual similarity between the query image and a previously-processed training image and tend to work best when landmarks or other unique features are visible. In the general case, accurately localizing images can be challenging. Even with the assistance of a search engine, users not only need to identify visual clues, but understand them well enough to translate into a search query. 
 Because geolocation is a task that requires investigation, in that analysts must collect sometimes disparate clues to uncover the origin of the image, it can be expected that users will formulate multiple queries as they seek to retrieve information about these clues. This task takes advantage of both the lookup abilities of a search engine and the contextual knowledge from humans, making it a compelling task to evaluate how users adapt their query formulation strategies. 

We conducted a between-subjects study with 60 participants  
randomly assigned to use either traditional or LLM-based search to
aid in image geolocation in order to address the following research questions: 
\begin{description}
 \item[RQ1] How does the use of an LLM-based search tool versus a traditional search tool impact participants' performance in geolocation tasks?
 \item[RQ2] How do participants adapt their query formulation 
 strategies when using LLM-based search compared to traditional search for image geolocation?
\item[RQ3] What are the key challenges encountered by participants when using LLM-based search for image geolocation?
\end{description} 
Our results indicate that participants using traditional search outperformed those using LLM-based search in terms of accurate image geolocation. This outcome can be explained by our qualitative findings, where participants reported challenges formulating queries when interacting with the LLM-based search engine. LLM-based search users issued longer, more conversational queries within shorter search sessions. Participants using the traditional search engine tended to extend their initial queries with additional terms when reformulating, while those utilizing the LLM-based search consistently rephrased their initial queries.

\section{Related Work}

Search engines have evolved into indispensable tools that influence how information is accessed and problems are solved~\cite{haider2019invisible}. However, effectively communicating the user's search intent has been a persistent challenge. Much work has been dedicated toward understanding web search query formulation patterns~\cite{jansen2006we,pang2011search,10.5555/1613715.1613848} and investigating how users adapt their queries and reformulation strategies in efforts to uncover search intent~\cite{10.1145/2600428.2609614, 10.1145/3442381.3450127,SearchSuccesses2017}. 
These strategies can be domain-specific. For instance, for health-related information, Zuccon showed that search results were less helpful when users issued complex queries describing their symptoms rather than using medical terminology~\cite{zuccon2015diagnose}. In the educational setting, students heavily rely on search engines for academic purposes~\cite{jadhav2011significant, salehi2018use}. However, it has been shown that a substantial portion of academic search sessions result in null queries, when individuals use vague or complex terms resulting in empty search results and obstructing users from achieving their intended search objectives~\cite{searchfailures2017}. 
Recognizing and understanding these challenges related to user behavior and query formulation strategies can enhance the overall search and retrieval experience. Our work builds upon existing research in web search and query formulation and extends the analysis to LLM-based search for the task of image geolocation. 

\subsection{LLM-based Search Analysis} 

LLMs are trained on large amounts of text corpra, and their effectiveness in various applications hinges on the ability to query them effectively~\cite{adolphs2021query}.
In efforts to optimize LLMs for retrieval tasks, several works have investigated the process of querying LLMs for specific information. Jiang~\cite{jiang2020know} highlighted the consequences of poorly written prompts, yielding failed retrieval results and proposed a method that used multiple automated paraphrases of the query and an aggregation scheme, mirroring how humans often rephrase their queries and provide additional context to make them more informative. Similarly, the work of Petroni~\cite{petroni2020context} examined enhancing the LLM retrieval by augmenting queries with relevant context and demonstrated improved performance on various LLMs on factuality tests. 

When humans seek information, they often clarify their queries with examples to obtain better results. To mimic this behavior, Brown employed few-shot learning, which involves conditioning the LLM on the task description and just a few examples, and found that this  ``in-context learning'' works best with larger language models~\cite{brown2020language}. A more recent effort argued that language models do not learn tasks during runtime from few-shot examples, but locate tasks within the model's preexisting knowledge; this paper proposes 0-shot prompts, which uses an alternative query with different phrasings to provide additional task descriptions~\cite{reynolds2021prompt}. Wei introduces chain-of-thought, which aims to replicate the human thought process when addressing complex problems~\cite{NEURIPS2022_9d560961}.  These efforts aimed at enhancing the LLM retrieval, but do not address 
the challenges faced by non-expert users when querying LLMs. Recent work~\cite{10.1145/3544548.3581388} involves non-experts issuing prompts for LLM-based chatbots and found that struggles in formulating effective prompts resemble issues observed in end-user programming and interactive machine learning systems. Their work emphasizes the need for further research in LLMs and prompt-literacy, specifically for non-expert users. Our work focuses on this challenge for the multifaceted task of image geolocation.   

\subsection{Image Geolocation}

Image geolocation is a widely-studied task. One effort used a carefully constructed dataset to investigate the types of clues and strategies users employ for image geolocation~\cite{mehta2016exploratory}. Several efforts have addressed the labor-intensive nature of image geolocation
by incorporating crowdsourcing to improve location identification. One study introduced a
diagramming technique involving visual representations from a bird's-eye or satellite perspective, which allowed novice crowd workers to 
collaborate with experts~\cite{Kohler_Purviance_Luther_2017}. In a follow-up study, the authors introduced GroundTruth, a system that enhanced 
image geolocation accuracy through shared representations for crowd-augmented expert work~\cite{venkatagiri2019groundtruth}.

Other studies have explored how to improve the accuracy of non-expert workers in image geolocation tasks. 
One project explicitly instructed novice users to follow a three-step workflow inferred from expert strategies:
collecting image-related clues, deriving potential coordinates based on these clues, and identifying the image 
location on a map~\cite{kim2022}.  Another method~\cite{CROWD4EMS2019} introduced a crowdsourcing 
platform that leverages existing data mining methods to estimate photo and video locations from social media, 
then used crowdsourcing for verification. 

In our approach, we focus on how participants articulate visual clues into search queries, 
and whether those query strategies differ depending on the type of search tool available.

\begin{figure*}
    \centering
    \begin{subfigure}[b]{.32\textwidth}
        \centering
        \includegraphics[width=\linewidth]{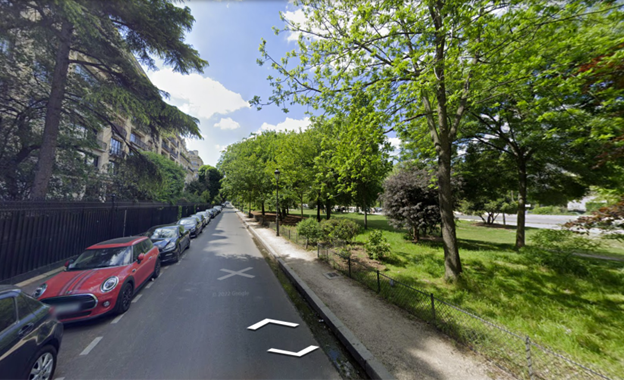} 
        \subfloat{Paris, France}
    \end{subfigure}
    \begin{subfigure}[b]{.32\textwidth}
        \centering
        \includegraphics[width=\linewidth]{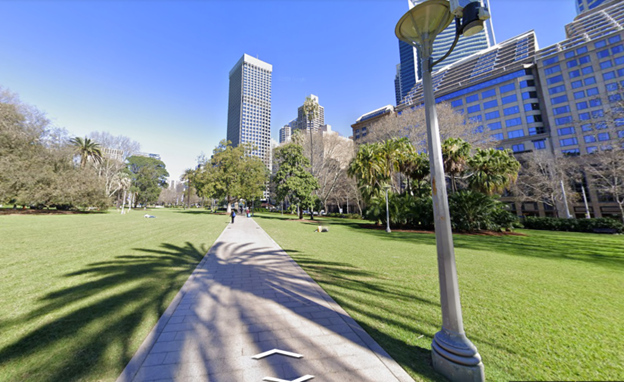}
        \subfloat{Sydney, Australia}
    \end{subfigure}
    \begin{subfigure}[b]{.32\textwidth}
        \centering
        \includegraphics[width=\linewidth]{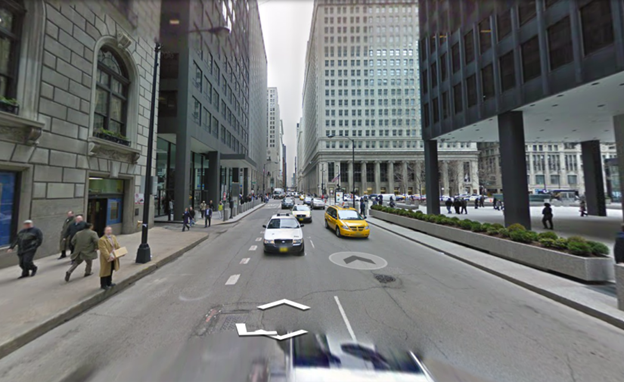} 
        \subfloat{Chicago, USA}
    \end{subfigure}
    \\
    \begin{subfigure}[b]{.32\textwidth}
        \centering
        \includegraphics[width=\linewidth]{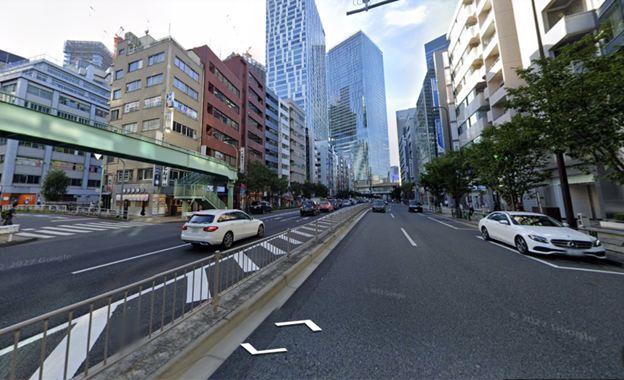}
        \subfloat{Tokyo, Japan}
    \end{subfigure}
    \begin{subfigure}[b]{.32\textwidth}
        \centering
        \includegraphics[width=\linewidth]{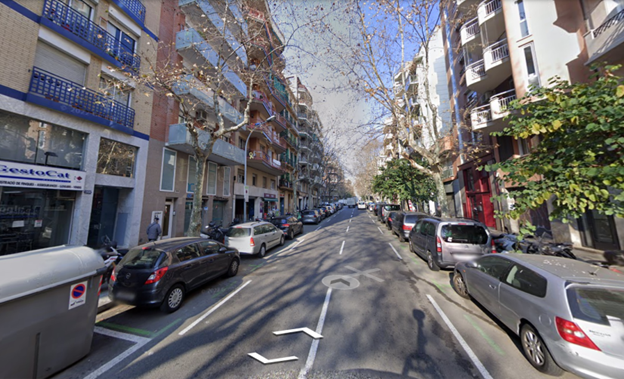}
        \subfloat{Barcelona, Spain}
    \end{subfigure}
    \begin{subfigure}[b]{.32\textwidth}
        \centering
        \includegraphics[width=\linewidth]{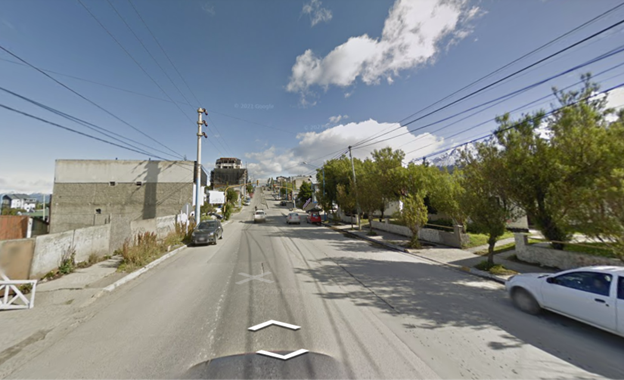}
        \subfloat{Ushuaia, Argentina }
    \end{subfigure}
    
    \caption{Initial viewpoints (with the location indicated) of the six rounds in the experiment.}
    \label{fig:geoguesser-locations}
\end{figure*}

\section{Methods}
We conducted a between-subjects user study involving 60 participants. In this section, we describe the experimental platform, 
recruitment of participants, task design, and measures.

\subsection{Experimental Platform}

Image geolocation has been well-studied due, in part, to the popularity of gamified versions of the task. 
The most well-known version is GeoGuessr; others include GeoGuess, Geotastic, and City Guesser. 
The objective of these games is to predict the correct 
location on map given an image, video, or other information and points are accumulated based on speed and/or accuracy. 
These games can serve as useful platforms for evaluating a wide variety of cognitive tasks.
In this study, we use GeoGuess~\cite{geoguess}, an open-source 
image geolocation game. Users are presented a Google Map StreetView image
 and have two minutes to guess the
location by dropping a pin on a world map. Users can navigate  
using the StreetView interface to virtually zoom and move through the scene. Up to 5,000 points can be earned based on proximity of the prediction to the actual location.
In the instructions for the game, users are encouraged to seek out clues that may be useful for localization.

\subsection{Participants}
Participants, 18 years of age or older who could read and understand English, were recruited on a university campus.  
Our sample consisted of 60 participants 
whose mean age was 25 (\textit{SD} = 4.95) with a diverse range of academic backgrounds, majoring in accounting, biology, 
computer science, systems engineering, physics, chemistry, global studies, and marketing. The IRB approved study was carried out by 
two members of the research team over the span of three weeks. All participants received a \$5 gift card to a coffee shop 
and were informed of the task description, 
duration, compensation, and their right to forfeit at anytime before participating.

\subsection{Task Design} 

Users were provided a dual-monitor setup, with the geolocation task on one screen 
and the search engine on the other. Microsoft Bing served as the traditional search engine and 
Microsoft Bing Chat, which is powered by ChatGPT, served as the LLM-based search engine.  

We followed a between-subjects study design, where each participant was randomly assigned to either the (Traditional) 
Search or the LLM condition. 
The experiment consisted of six geolocation tasks (shown in Figure~\ref{fig:geoguesser-locations}), which were intended to vary in difficulty. 
Participants were provided instructions to only use the provided search engine and not perform image-based search. 
Participants were asked to confirm their understanding of the instructions by clicking on a (I understand) button.
After the instructions, participants watched a short instructional video on how to use the geolocation interface. 

For each round, the participant had two minutes to provide a guess. They could consult the search tool as
often as they needed, given the time constraint. Upon completion of the six rounds, 
participants were invited to fill out a post-study survey with questions about 
familiarity with image geolocation, traditional or LLM-based search, attitudes toward artificial intelligence, 
and a set of open-ended questions for additional feedback. The entire experiment was designed to be completed in approximately 15 minutes per participant.
 
\subsection{Measures}

The primary dependent variable in this study is performance, measured by the points earned by each participant per round. 
The score ranged from 0 to 5000; the maximum score was obtained when the prediction was within a few kilometers of the actual location.
The primary independent variable, type of search, was modeled as a fixed effect in our linear mixed-effects model. Each participant played the six rounds in the same order. The round number, which correlated with difficulty, was modeled as a random effect. 
For each participant, we maintained an event log of timestamped actions that included switching between web search and geolocation. Additionally, we recorded the search queries. 

\begin{table}
\centering
\caption{Geolocation performance on the six round experiment.}
\label{table:lmm-fixed-effects}
\begin{tabular}{p{2cm}cccc}
\hline
& Estimate & Std. Error & t-value & p-value \\
\hline
(Intercept) & 2678.5 & 167.4 & 15.999 & <2e-16 *** \\
Condition(Search) & 501.3 & 243.5 & 2.059 & 0.0414 * \\
\hline
\end{tabular}
\label{tab:LMM-performance}
\end{table}

\begin{figure*}
  \centering
  \includegraphics[width=\linewidth]{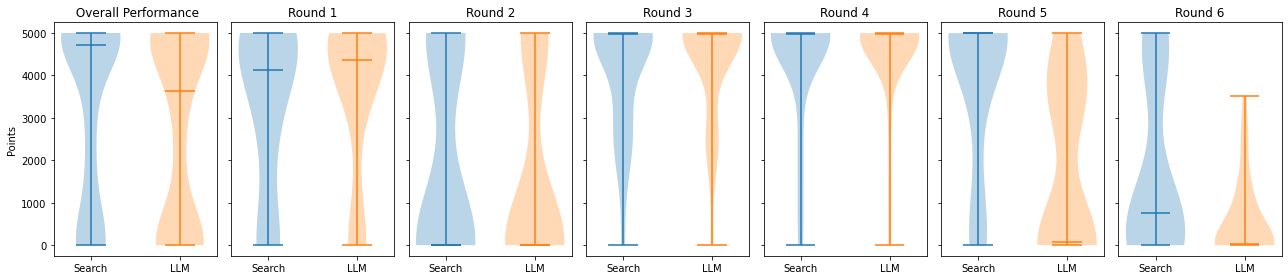}
  \caption{Performance distribution (points) comparison between Search and LLM conditions on average (left) and per round.}
  \label{fig:performance-distribution}
\end{figure*}

\section{Analysis \& Results}
We excluded the five participants that did not engage with the 
search engine for any of the rounds, which left 29 for the Search condition and 26 for the LLM condition. 

For the Search condition, the mean performance score was 3189, with a median of 4712 and an IQR of 4262.25. 
For the LLM condition, the mean performance score was 2725, with a median of 3637.5 and an IQR of 4952. 
Figure~\ref{fig:performance-distribution} shows the distribution of scores across the two conditions by average and across the six rounds.

Overall, participants in the Search condition outperformed those in the LLM condition. 
For rounds 1-4, the performance was similar for both conditions, with both groups finding round 2 challenging. 
In the final two rounds, participants across both conditions performed poorly, with those in the LLM group performing notably worse.
We use a Linear mixed-effects Model (LMM) to evaluate the difference in performance between the groups. The results are shown
in Table~\ref{tab:LMM-performance}. There was a significant difference in performance between the two conditions ($p = 0.0414$).

\subsection{Query Formulation Patterns}
\label{sec:query-formulation}

Query formulation is a fundamental tool in the analysis of search behavior. Here, 
we investigate four key query formulation metrics, comparing their differences across the two conditions.

\begin{figure}
  \centering
  \includegraphics[width=.75\linewidth]{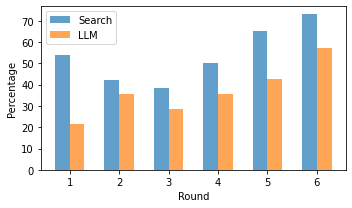}
  \caption{Percentage of multi-query rounds for Search and LLM conditions}
\label{fig:sq-precentage}
\end{figure}

\subsubsection{Number of Queries}
\label{tab:session-length}
We examined the number of queries per round. Participants 
in the Search condition issued an average of 1.98 queries, whereas those in the LLM condition 
issued an average of 1.04 queries. A Chi-Square test showed this association to be
significant ($\chi^2(6) = 19.71, p = 0.003$); users in the Search condition issued more queries.

We computed the percentage of rounds in which participants issued more than one query. As shown in Figure~\ref{fig:sq-precentage},
participants in both conditions issued more queries as the task increased in difficulty.
Participants in the Search condition favored issuing more queries, starting at around 55\% and reaching 70\% by the end of the task. In the LLM condition, participants issued more queries at a lower rate starting at 20\%, but reaching 55\% by the last round.

\subsubsection{Query Length}

Query length, the average number of terms in each query, can provide valuable insights into query formulation patterns~\cite{10.1145/1935826.1935842}.  
On average, participants in the Search condition formulated queries comprising 4.19 terms. 
In contrast, participants in the LLM condition issued queries with an average of 6.06 terms. A Chi-Square test revealed a highly significant association between the conditions and the differences in query length patterns ($\chi^2(28) = 61.78, p < 0.001$).

\begin{figure}
  \centering
  \includegraphics[width=.85\linewidth]{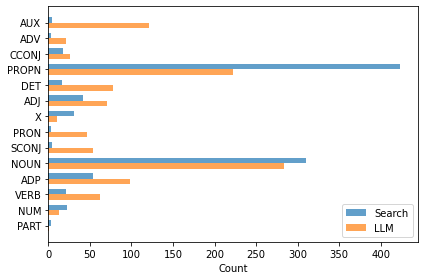}
  \caption{Comparison of part-of-speech tag counts between Search and LLM conditions }
\label{fig:pos}
\end{figure}

\subsubsection{Part-of-Speech Tagging}
To explore potential differences in linguistic characteristics, we performed part-of-speech tagging. 
Figure~\ref{fig:pos} shows the distribution of tags in the queries across conditions. 
After adjusting the alpha value using Bonferroni correction, in the LLM condition, several tags, including adverbs (ADV), adpositions (ADP), determiners (DET), auxiliary verbs (AUX), and (VERB) 
exhibited significantly higher frequencies ($p = 0.051$, $p = 0.046$, $p <0.001$, $p <0.001$, $p =0.019$) than the Search condition. 
The increased use of adverbs and auxiliary verbs in the LLM queries suggested a more natural language style, potentially influenced 
by the conversational nature of interactions with LLMs~\cite{10.1145/3406522.3446035}. On the other hand, in the Search condition, 
usage of proper nouns (PROPN) was significantly higher ($p = 0.034$) than in the LLM condition. This indicates a greater tendency 
to perform a keyword-based search using specific entities or locations when interacting with a traditional search engine.

\subsubsection{Questions}
We explore the categorization of \emph{question} and \emph{non-question} queries. Following Pang and Kumar~\cite{pang2011search}, we defined question queries based on the following criteria: 
\begin{itemize}
    \item Interrogative start: Queries that start with \emph{how, what, which, why, where, when, who, whose}.
    \item Modal verb start: Queries that start with \emph{do, does, did, can, could, has, have, is, was, are, were, should}. However, an exception is made for queries where the second word is \emph{not}.
    \item Queries that end with a question mark (?).
\end{itemize}
Queries not meeting the criteria were classified as \emph{non-question}. 

For the Search condition, only 17\% were question queries. Conversely, for the LLM condition, 73\%, were question queries. A Chi-square analysis yielded statistically significant results between the two types ($\chi^2(1) = 6.37, p = 0.012$). These findings suggest that participants in the LLM condition applied a more conversational style. 

\subsection{Query Reformulation Strategies}
\label{sec:query-reformulation}

We explore how users reformulate and refine their queries during a given round, focusing on two primary aspects: changes in query length and term repeats, which allows us to understand how participants progressively adapt their queries. 

\begin{figure}
    \includegraphics[width=.75\linewidth]{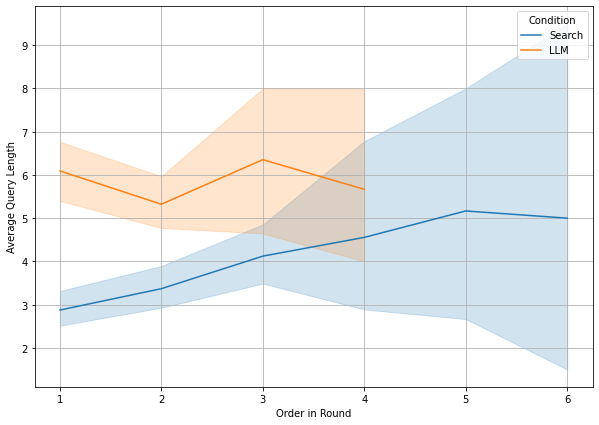}
    \caption{Average number of query terms for successive queries in a round}
\label{fig:query-length-pos}
\end{figure}

\subsubsection{Number of Terms}

We investigate how the length of queries changes within round. 
For each round, we computed the average number of terms in each query in order. Figure~\ref{fig:query-length-pos} shows the average number of
terms by query order. Participants in the LLM condition issue an initial query
of $\sim 6$ terms and maintain this length for subsequent queries. Meanwhile, participants in the Search condition tend to start with shorter ($\sim 3$) queries and gradually increase. LLM users, favoring longer queries, may indicate a tendency for conversational interactions. Conversely, participants in the Search condition, may reflect an initial focus on keyword-driven retrieval, with subsequent query expansion.

\begin{figure}
  \centering
  \includegraphics[width=.75\linewidth]{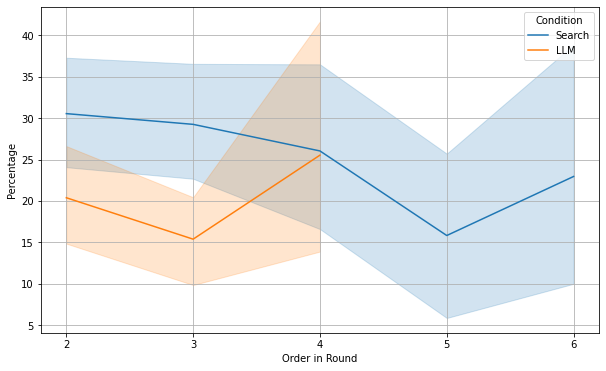}
  \caption{Percentages of term repeats for successive queries in a round}
\label{fig:query-term-reuse}
\end{figure}

\subsubsection{Term Repeats} 
We examine term repeats within a round to understand how often users refined their 
initial queries. We computed the Jaccard similarity percentages~\cite{ExplorationQR} of consecutive queries in a round. Figure~\ref{fig:query-term-reuse} shows percentages of queries that share identical terms with the previous query in a round. Initially, participants in the LLM condition had generally lower term reuse of around 20\%, suggesting a moderate level of query refinement. In the Search condition, participants began with a higher term reuse rate of 30\% with a gradual decline as the round progressed, which suggests that participants initially focused on refining their queries, then shifted to queries formulated differently or focused on new clues.

\begin{figure}
  \centering
  \includegraphics[width=.99\linewidth]{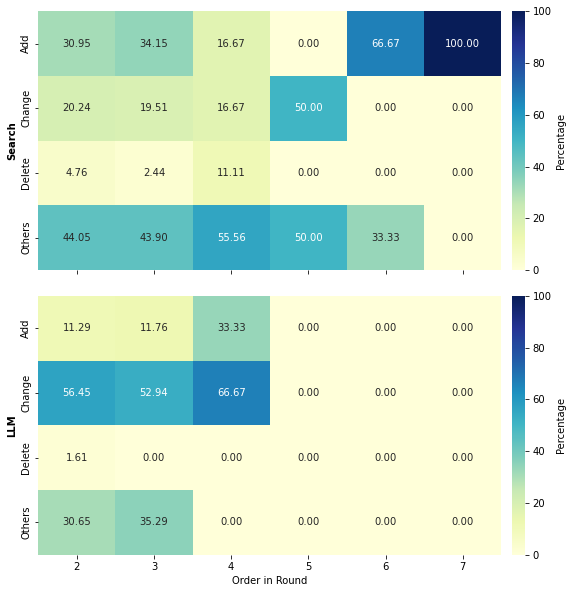}
  \caption{Distribution of syntactic-level types across ordered queries for Search (top), and LLM (bottom) conditions}
\label{fig:syntactic-combined}
\end{figure}

\subsection{Query Reformulation Types}

Analyzing query reformulation types (QRTs) allows us to infer the user intent in query reformulation. 
We adopt the QRT taxonomy proposed by Chen et al.~\cite{10.1145/3442381.3450127}, which characterizes QRTs at both the syntactic and intent level.

Syntactic changes in consecutive queries, which involve alterations in the structure and composition, are categorized into five types:
\begin{itemize}
\item \textbf{Add}: New terms are introduced into the query, resulting in an expansion of its content.
\item \textbf{Delete}: Terms present in the previous query are removed in the current query.
\item \textbf{Change}: Modifications involve replacing some terms while keeping others unchanged.
\item \textbf{Repeat}: A query remains identical to the preceding one.
\item \textbf{Others}: A combination of different changes within a query or the introduction of an entirely new query.
\end{itemize}

We compute the percent of each syntactic category type for ordered queries in a round.
Figure~\ref{fig:syntactic-combined} shows the distribution of syntactic-level QRTs for the first, second, etc. query issued in each round. For the Search participants (Figure~\ref{fig:syntactic-combined} (top)), we observe that the predominant QRTs were ``Add'', ``Others'', and ``Change.'' The high initial rate for ``Other'' suggests an exploratory intent at the onset. As the task progressed, there was a noticeable rise in the ``Add'' type, indicating adding details to their queries as they solved the task, suggesting more exploitative behavior.

For the LLM condition (Figure~\ref{fig:syntactic-combined} (bottom)), the most frequent QRT was ``Change'', showing a steady increase as participants progressed. Compared to the Search condition, the ``Add'' type was less common. This lower 
occurrence of ``Add'' suggests that participants in the LLM condition were less inclined to augment their queries with additional terms.

\begin{table}
\centering
\resizebox{\columnwidth}{!}{%
\begin{tabular}{p{1.5cm}|p{4cm}|p{4cm}}
\hline
\multicolumn{1}{c|}{\textbf{Intent Category}} & \multicolumn{1}{c|}{\textbf{Definition}} & \multicolumn{1}{c}{\textbf{Example Query}} \\
\hline
Specification (Spec) & Query becomes more specific, narrowing down the search intent & sauf street sign $\rightarrow$ sauf street sign handicap red x \\
\hline
Generalization (Gen) & Query becomes more general, broadening the search intent & saint james peter adam Hamilton $\rightarrow$ saint james \\
\hline
Synonym (Syn)& Substitution of a term with its synonym while maintaining the overall meaning &RUE CREVAUX street MAP $\rightarrow$ RUE CREVAUX street location \\
\hline
Somewhat Relevant (SR)& Intent shifts slightly while remaining  somewhat tied to the original query & what language is SAUF in? $\rightarrow$ how about rue crevaux? \\
\hline
New Topic (New)& Intent shifts significantly to a different subject & what countries have placacentro masisa? $\rightarrow$ where is 17 de mayo in Chile? \\
\hline
Others (Oth) & Queries that do not fit any category & federal street boston $\rightarrow$ federal street in (Boston) \\
\hline
\end{tabular}%
}
\caption{Intent-level query reformulation types}
\label{tab:intent-level-categories}
\end{table}

Beyond the syntactic level, we examined the intent level QRT. Rather than measuring 
\emph{how} users modify queries, the aim is to measure \emph{why} the changes were made to uncover the underlying motivations, evolving information needs, and user goals. Table~\ref{tab:intent-level-categories} introduces the six intent categories along with example queries from our dataset.

\begin{figure}
  \centering
  \includegraphics[width=.99\linewidth]{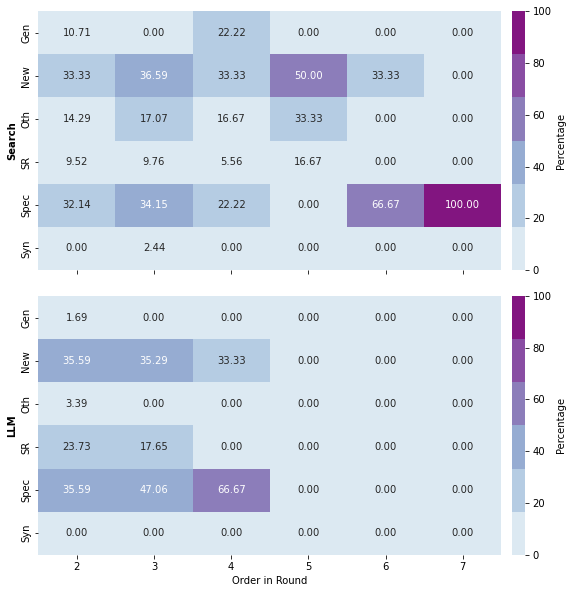}
  \caption{Distribution of intent-level types across ordered queries for Search (top), and LLM (bottom) conditions}
\label{fig:intent-combined}
\end{figure}

Two members of the research team performed intent-level categorization. 
Both researchers individually categorized the queries, then met
to reach a consensus on any discrepancies. As shown in (Figure~\ref{fig:intent-combined} (top)), for participants in the Search condition ``Specification'', and ``New Topic'' were the predominant QRTs.
Comparing these findings with the syntactic changes observed in (Figure~\ref{fig:syntactic-combined} (top)), we notice a similar pattern between ``Specification'' and ``Add''. This pattern suggests that participants were narrowing down their search intent by adding more details to their query. A parallel trend is observed between ``Others'' in the syntactic types and ``New Topic'' as their percentages initially fluctuate but eventually follow a similar pattern. 

For the LLM participants, the distribution of intent-level QRTs is shown in (Figure~\ref{fig:intent-combined} (bottom)). ``Specification'' was the dominant category. Analysing similar trends with the syntactic changes shown in (Figure~\ref{fig:syntactic-combined} (bottom)), there was a steady increase in both `Specification'' and ``Add'', however ``Specification'' being much more frequent than ``Add.'' Interestingly, a parallel trend can be observed between ``Specification'' and ``Change.'' This suggests that while participants using the LLM primarily focused on narrowing down their search intent, they did so without necessarily adding terms to their queries. This behavior differs from the Search condition trend, where both ``Specification'' and ``Add'' showed increasing percentages and similar frequencies. This contrast showcases the distinctive user interactions between the two assistants. While Search condition ``Specification'' often involves query expansion, in the LLM condition, ``Specification'' primarily shows as query rephrasing.

 \section{Qualitative Findings}
 The qualitative findings derive mainly from the responses to the post-study survey and a comparison of the search results returned by each search engine for similar queries.

\subsection{Open-ended Survey Questions}
We conducted a post-study survey with open-ended questions 
to better understand how participants translated clues into search queries and the challenges they faced.

 \begin{figure}
  \centering
  \includegraphics[width=.95\linewidth]{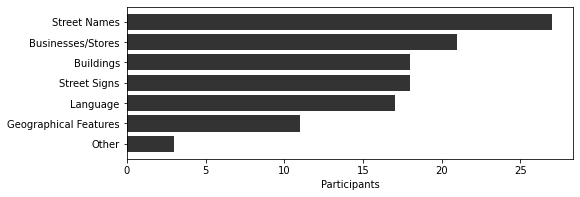}
  \caption{Top clues by participants across conditions}
\label{fig:clues}
\end{figure}

\subsubsection{Clues Identified by Participants}
We coded and categorized responses to the 
question: \textbf{What types of clues did you identify that helped you with the task?}. As illustrated in Figure~\ref{fig:clues}, the predominant clue category was street names, followed closely by business names, including store names. Language was also helpful to our participants, particularly to identify non-English speaking countries. Some participants also referred to geographical features like mountains and large bodies of water. The identified clues were consistent across all participants, regardless of the assigned search tool; this aligns with prior work that explored image geolocation~\cite{mehta2016exploratory}.

\begin{table*}
\centering
\begin{tabular}{p{3.6cm}|p{4.75cm}|p{6.25cm}}

\hline
\multicolumn{1}{c|}{\textbf{Qualitative Code}} & \multicolumn{1}{c|}{\textbf{Description}} & \multicolumn{1}{c}{\textbf{Example Response}} \\  
\hline
Location Specificity & Precisely specifying and differentiating streets within a city & P1(Search): I did have a hard time. There were many First and Second streets I was on, and it is difficult to distinguish those between the first and second streets of other cities \\
\hline
Language Barriers & Identifying Foreign Words with Non-English alphabets & P13(LLM): My strategy of finding word clues failed if the words I saw were in a language that does not use the English alphabet \\
\hline
Interpreting Visual Clues & Translating visual clues into effective questions or searches & P58(Search): It is hard to try to search the architecture of a building without using an image search \\
\hline
Efficient Query Formulation & Crafting efficient search queries that would yield precise results & P38(LLM): It was challenging to find the correct wording to get the desired result \\
\hline
Lack of Textual Clues & Locations with limited textual clues, for example remote locations & P53(LLM): Sometimes it's very hard to find street names or shop names from the image, especially if the images are from remote locations \\
\hline
Effective Clue Selection & Finding clues that can be described or will generate effective results when searched & P50(LLM): Figuring out what clue to look up, for example, local places were useful, meanwhile large chains are not as useful \\
\hline
\end{tabular}
\caption{Qualitative codes resultant from the coding of challenges described by participants, and example quotes}
\label{tab:qualitative-codes}
\end{table*}

\subsubsection{Translating Clues into Search Queries}
\label{sec:strategies}
The post-study survey asked: \textbf{How did you translate the clues into search queries?}. Using an inductive, open coding approach~\cite{thomas2006general}, we coded and categorized these responses into distinct strategies:

\paragraph{Language Identification} 10 participants from each condition focused on identifying the languages present on signs, buildings, and stores. P13 stated, ``I would type in the words I saw, and ask the helper, what language is this in?.'' Similarly, P18 explained, ``I translated some of the clues I saw to English, this way it shows me what the origin of the language.''

\paragraph{Locating Street Signs} 10 participants from the Search and 7 from the LLM condition utilized this strategy. They focused on finding street signs in corners and intersections to get closer to the location. P31 explained, ``I was searching for street names, trying to identify which neighborhoods the locations were in, for larger cities.'' P36 also said, ``I used the road signs to get a general idea of city and direction of city.'' 

\paragraph{Locating Businesses/Stores} Nine participants from each condition focused on locating businesses, stores, and shops. P37 mentioned, ``my strategy was typing company names into the helper.''

\paragraph{Describing Geographic Features} A few participants, mostly from the LLM condition, described the geographic features of the location. This included providing details about trees, architecture, and mountains. P13 explained, ``I tried describing the environment I was in to the chatbot, but the results were often not good.''

\paragraph{Locating Landmarks} In a similar approach to locating businesses, a small number of participants actively searched for large buildings and landmarks. P51 noted, ``I searched for a landmark around the area.''

Although these strategies did not differ significantly between the two conditions, analyzing how participants translated visual clues into search queries is important for gaining insights into their approach to image geolocation.

\subsubsection{Challenges Identified by Participants}
\label{sec:challenges}
We asked the participants to describe the challenges they faced during the experiment. Only a few participants did not answer or stated they faced no challenges. The responses were coded into in six qualitative classes. Each code, with an example participant response, is given in Table~\ref{tab:qualitative-codes}.

\begin{figure}
  \centering
  \includegraphics[width=.85\linewidth]{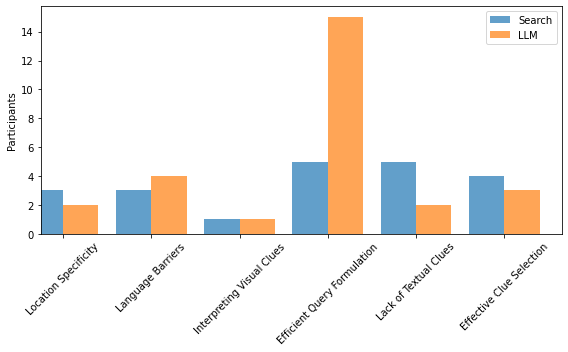}
  \caption{Comparison of challenges faced by our participants in both the Search and LLM conditions }
\label{fig:challenges}
\end{figure}

The distribution of challenges across the experimental conditions is shown in Figure~\ref{fig:challenges}. There is notable discrepancy for \emph{Efficient Query Formulation}. While a few participants from the Search condition did mention this challenge, it emerged as the primary obstacle for over half of the participants in the LLM condition. P5 explained, ``My biggest challenge was getting the chatbot to understand exactly what I wanted.'' Similarly, P17 said, ``Trying to be concise and precise with my searches using the chatbot was challenging.'' These participants encountered difficulties in effectively communicating their intent to the LLM-based search engine. Others took a more strategic approach to address this challenge. P46 explained, ``I realized I needed to ask less specific questions and go more broadly to get answers.'' This adaptive strategy reflects participants' attempts to optimize their interactions with the LLM. Participant P55 mentions, ``The challenging part was figuring out what I was looking at and translating it to a question that would narrow down answers coming from the chatbot.''

Another challenge worth highlighting was the language barrier. Despite the advanced language capabilities of modern LLMs, the participants had difficulties when formulating language-related questions. P49 stated, ``I got a response in Spanish when I typed a Spanish building name but the chatbot didn't answer the question I was asking.'' Similarly, P28 describes, ``I had to rephrase a couple of times, especially If I was asking about phrases in different language than English the chatbot would shoot back a failed search.'' 

\begin{figure}
    \begin{subfigure}[b]{0.463\textwidth}
        \centering
        \includegraphics[width=0.45\linewidth]{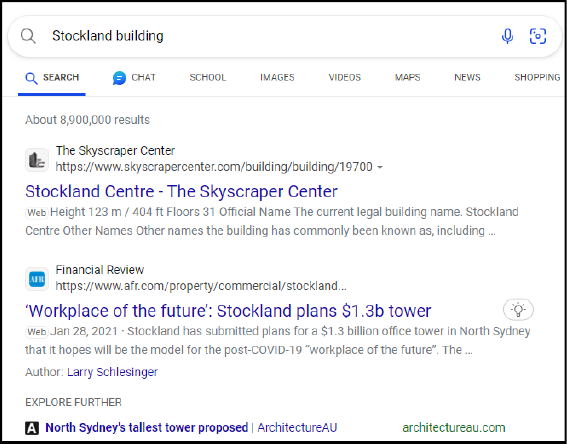}\hfill
        \includegraphics[width=0.55\linewidth]{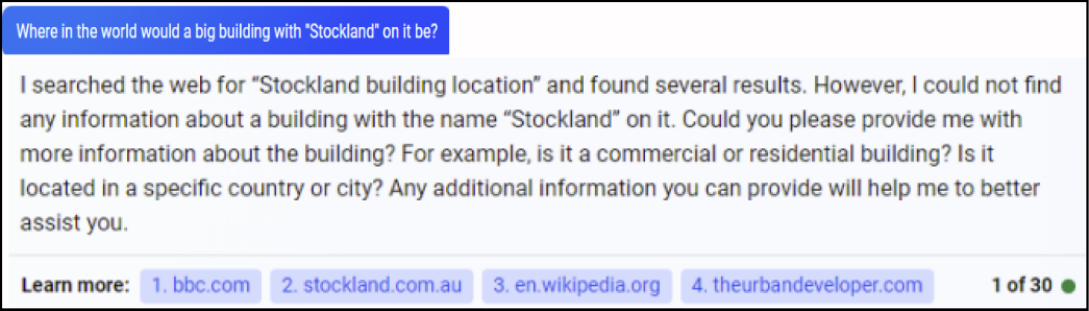}
        \caption{Round 2 clue: Stockland Building}
        \label{fig:round2-results}
    \end{subfigure}
     \begin{subfigure}[b]{.463\textwidth}
        \centering
        \includegraphics[width=0.45\linewidth]{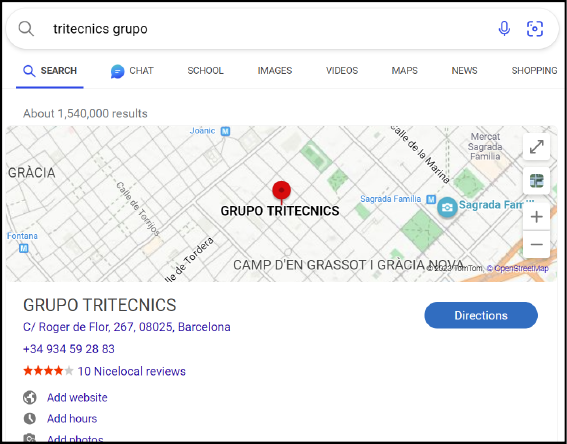}\hfill
        \includegraphics[width=0.55\linewidth]{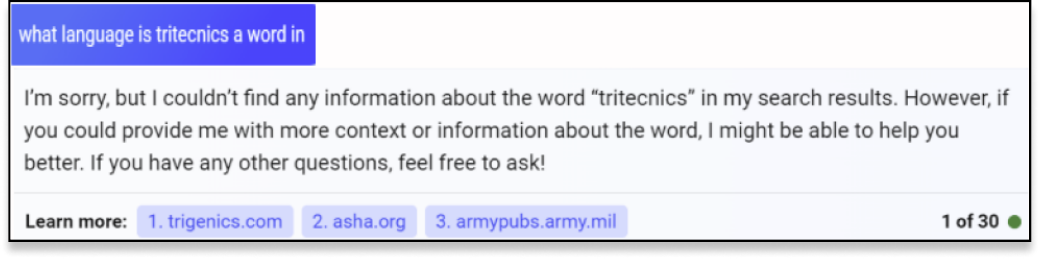}
        \caption{Round 5 clue: tritecnics}
        \label{fig:round5-results}
    \end{subfigure}
    \begin{subfigure}[b]{.463\textwidth}
        \centering
        \includegraphics[width=0.45\linewidth]{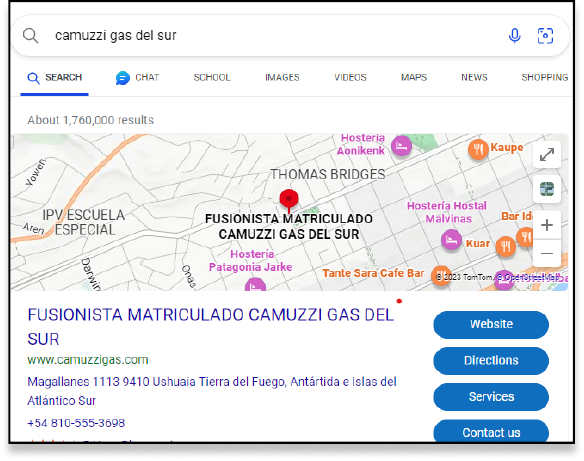}\hfill
        \includegraphics[width=0.55\linewidth]{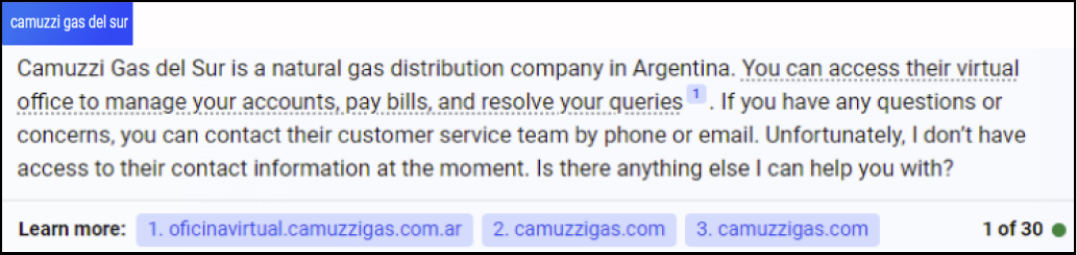}
        \caption{Round 6 clue: Camuzzi gas del sur}
        \label{fig:round6-results}
    \end{subfigure}
    
    \caption{Comparison of results obtained from asking about the same clue by participants using different search assistants}
    \label{fig:example-results}
\end{figure}

\subsection{Search Results}
\label{sec:se-vs-llm}
We examined the search queries and the results. For instance, as shown in Figure~\ref{fig:round2-results}, when participants searched for ``Stockland building'', LLM search did not return any results and requested clarification, while the search engine quickly located the building in Sydney, Australia, in the first few results. 

Similar disparities were observed when searching in languages 
other than English. For instance, when attempting to locate the ``tritecnics'' company, as shown in Figure~\ref{fig:round5-results}, the LLM-based search struggled to translate or make sense of the word, whereas the search engine produced map results for Barcelona, Spain. In Figure~\ref{fig:round6-results}, while the LLM-based search 
successfully replied that ''Camuzzi gas del sur'' was a gas distribution company in Argentina, the search engine identified the location as the remote city of Ushuaia and presented a map, effectively solving the task for those participants.

\section{Discussion}
The results demonstrate some key findings: (1) participants using  traditional search for assistance outperformed those using the LLM-based search, (2) distinct strategies emerged due to significant variations in the query formulation strategies between the two conditions, and (3) the qualitative findings revealed that participants using the LLM search struggled to effectively formulate their queries.

\subsection{Differences in Performance}
In response to \textbf{(RQ1)}, the results showed that participants using the traditional search outperformed those using the LLM search. We believe that a substantial portion of the performance difference can be explained by the difference in results for similar queries, as shown in Section~\ref{sec:se-vs-llm}.
Specifically, as shown in Figure~\ref{fig:round5-results} and Figure~\ref{fig:round6-results}, the search engine provided a map, effectively directing our participants to the exact location. LLM-based search should not only provide creative answers, but the same  features as a traditional search engine, including maps. As evident from the examples in our study, LLM search did not always return results that should have been within its capabilities. Perhaps, the participants needed to explicitly request maps or specific directions to effectively utilize these features; however this was not necessary for traditional search. This highlights the challenge of crafting effective prompts, which, as research has shown~\cite{10.1145/3544548.3581388}, significantly influences the output of LLM-based methods. While ``Learn more'' links were included with each answer, we found that none of the participants took advantage of this feature. This raises questions about the perceived affordances of LLMs compared to traditional search engines, as the integration of similar features in LLMs may not be as intuitive, as our study suggests. This underutilization of help links in LLM responses highlights the challenges of transitioning from traditional search engines to LLM-based search and aligns with Gibson's theory~\cite{gibson2014ecological} emphasizing the interaction between users and their technological environment, which shapes the possibilities and constraints for action.

In response to~\textbf{(RQ3)}, our qualitative analysis provided valuable insights into the performance disparities. Over half of the participants using the LLM-based search expressed difficulties in formulating their queries. Some expressed hesitation, while others found it challenging to form queries that effectively communicated their information needs. Participants also struggled to formulate queries in different languages. Despite stated support of LLM search for many languages, there were reported instances of failed results. This emphasizes the challenge of effectively prompting the LLM to comprehend and respond to queries in diverse languages. These challenges were less prevalent when using the search engine. These observations align with our query formulation analysis. In Section~\ref{tab:session-length}, we noted that the average number of queries for LLM participants consisted of a single query. Figure~\ref{fig:sq-precentage} showed that LLM participants were less inclined to reformulate or issue more queries throughout the task compared to Search participants.

\begin{table}
\centering
\small
\resizebox{\columnwidth}{!}{%
\begin{tabular}{p{3.2cm}|p{6.2cm}|p{0.7cm}}
\hline
\multicolumn{1}{c|}{\textbf{Query}} & \multicolumn{1}{c|}{\textbf{LLM Response}} & \multicolumn{1}{c}{\textbf{Score}}\\
\hline
countries in southern hemisphere with mountains & The Southern Hemisphere is home to many mountain ranges, including the Andes, the Sierra Nevada de Santa Marta, and the Cordillera Oriental and Cordillera Occidental in South America. The highest peak in the Americas, Aconcagua, is also located in the Andes. In Africa, Mount Kilimanjaro is the tallest mountain and one of the world's largest volcanoes. The Rwenzori range of mountains in Africa were created during the formation of the Great Rift Valley. New Zealand's South Island has more than 24 mountains with elevations of 9,000 feet or more¹. I hope this helps! & 605 \\
\hline
patas para arriba location & (Original Spanish) La expresión “patas para arriba” se utiliza para describir una situación en la que todo está desordenado o confuso. ¿Hay algo más en lo que pueda ayudarte? (Translated) The expression “upside down” is used to describe a situation in which everything is disordered or confusing. Is there anything else I can help you with? & 1240 \\
\hline
24seventaxi location & I’m sorry but I couldn’t find any information about 24seventaxi location. Could you please provide me with more details about what you are looking for? & 1773 \\
\hline
I need to know where the street H. de Magallanes is located in the world & I’m sorry but I couldn’t find any information about H. de Magallanes street. Could you please provide more details about the location or any other information that might help me find it? & 18 \\
\hline
\end{tabular}%
}
\caption{Examples where participants developed inaccurate mental models, resulting in the formulation of single, poor-quality queries, which led to low performance}

\label{tab:example-llm-fail}
\end{table}

While it has been shown that individuals can quickly build mental models when interacting with LLM chatbots~\cite{hwang2021ideabot}, the quality of these models remains uncertain, especially considering the relatively new nature of LLM technology. Participants may not have had 
the experience to develop accurate mental models of LLM capabilities. In the absence of well-defined mental models, users struggle to predict outcomes or make sense of their interactions with LLMs, leading to instances where users pose vague and poorly defined queries while expecting the LLM to respond appropriately~\cite{10.1145/3544548.3581388, subramonyam2023bridging}. Table~\ref{tab:example-llm-fail} demonstrates a few examples of this behavior. It is plausible that these mental models influenced the participants to attempt only a single query, potentially resulting in worse performance. 

\begin{figure}
   \vspace{-10pt}
    \centering
    \includegraphics[width=.99\linewidth]{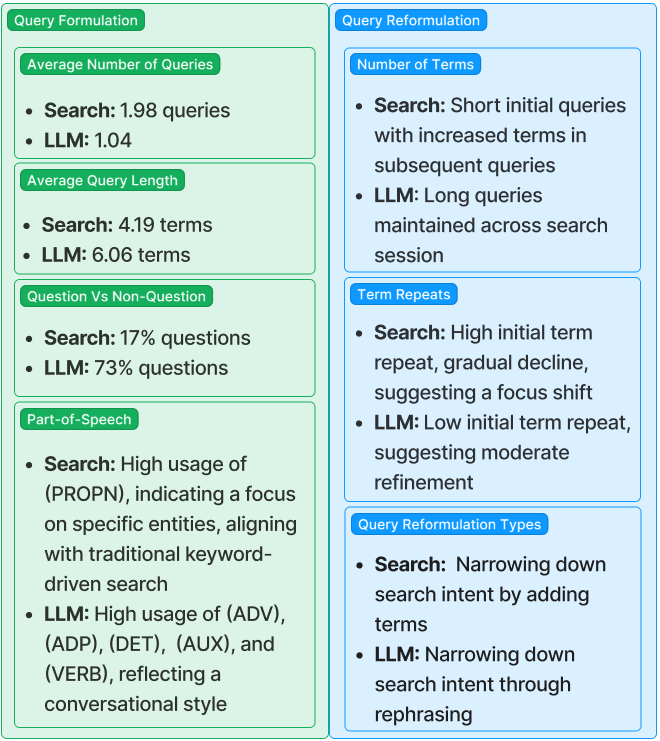}
    \caption{Summary findings of query formulation and reformulation analysis.}
    \label{fig:summary}
\end{figure}

\subsection{Differences in Query Formulation \& Reformulation Strategies}

Figure~\ref{fig:summary} presents a summary of our findings into query formulation and reformulation strategies, which is directly related to~\textbf{(RQ2)}. As described in Section~\ref{sec:query-formulation}, our study showed significant differences across all four formulation metrics. 
Participants in the Search condition issued shorter queries and a significantly increased use of proper nouns including places, store names, and streets. These findings align with prior research~\cite{jansen2006we,lewandowski2008search}, indicating that individuals are accustomed to keyword-based search from a lifetime of experience using traditional search engines.  In contrast, participants using the LLM-based search were inclined to issue longer and more natural language queries. These findings suggest that they adhered to the perceived norms of a conversational user interface~\cite{GoodConversation2019}. 

Several research efforts have explored the topic of query reformulation for traditional web search~\cite{aula2003query, 10.1145/2600428.2609614, 10.1145/1645953.1645966}. In our study, we adopted an existing taxonomy for categorizing both the syntactic and intent-based types of query reformulation, detailed in Section~\ref{sec:query-reformulation}. Our investigation revealed a notable trend. Although ``Specification'' or the act of narrowing down the scope and intent of a search, was the dominant category across both conditions, it aligned with different syntactic categories. In the Search condition, participants often expanded their queries by adding terms, a common behavior observed in prior research~\cite{chen2019investigating, 10.1145/3442381.3450127}. However, participants using the LLM search frequently narrowed down their search intents by paraphrasing their initial queries. This observation prompts a critical question of whether the development of new query reformulation taxonomies specific to LLMs could provide a better framework for understanding and characterizing user behavior.

\subsection{Geolocation Sensemaking Strategies}

As described in Section~\ref{sec:strategies}, participants employed diverse strategies when searching. In geolocation tasks, participants make sense of the clues within the images by using \emph{internal} or \emph{external} knowledge representations~\cite{mehta2016exploratory}. Our focus was on external knowledge representations cultivated through their searching. Therefore, we did not ask about internal knowledge, such as participants' cultural backgrounds or travel history.

In geolocation tasks, participants engage in sensemaking to interpret visual cues within images. Participants demonstrate adaptability in their approach, drawing from Pirolli and Card's sensemaking model~\cite{pirolli2005sensemaking}, which encompasses both top-down and bottom-up approaches. 
The top-down approach involves initiating the process with a theory or a broader concept and then seeking data to substantiate it. In this context, an example of the top-down approach is the strategy of language identification. Participants effectively employ this strategy by querying the assistant about the language's origin, which narrows the search scope to specific global regions. Similarly, some participants employed sensory sensemaking as a top-down approach by describing sensory aspects of the location to the assistant. However, the limited effectiveness of this strategy suggests that relying solely on sensory cues may not be sufficient for precise geolocation. The bottom-up approach, on the other hand, entails gathering data first and progressively forming a theory based on the available information. Examples of this approach included using street signs, buildings, and landmarks as reference points, facilitating the identification of cities then neighborhoods. 
The strategies identified in this study provide insights into user behavior and align with the cognitive processes driving geolocation sensemaking described by prior work~\cite{venkatagiri2019groundtruth}. Understanding how participants construct mental models based on image clues and apply sensemaking processes, enhances our ability to provide effective support and guidance in geolocation tasks.

\subsection{Limitations}
Amongst the study limitations was the latency in LLM-based search. Although the latency was only a few seconds, it could have disrupted the conversational flow, potentially affecting satisfaction and engagement during the task. Our study included participants with diverse backgrounds, education levels, ages, and degrees of technical literacy. This inherent diversity may have influenced how participants interacted with both the image geolocation task and the search engines. Furthermore, it's essential to acknowledge that image geolocation tasks have historically been conducted by expert image analysts. In our study, we did not explicitly categorize participants based on their levels of expertise in geolocation.  Lastly, our evaluation did not include the assessment of specific metrics such as search engine result pages (SERPs), clicks, or other performance indicators that could offer a more comprehensive view of the effectiveness of LLM-based search in image geolocation tasks.

\section{Conclusion and Future Work}

This study offered valuable insights into differences
in strategies and user behaviors when using traditional compared to LLM-based search for image geolocation. 
We examined the differences in performance, query formulation, and the sensemaking strategies employed by participants 
in these two conditions. Despite the growing capabilities of LLMs, the 
results reveal that participants using traditional search engines outperformed those relying on LLM-based search. 
An in-depth exploration of the distinct query formulation strategies 
utilized by participants mostly explained the performance difference, as evidenced by our qualitative findings, with query formulation being identified as the most challenging aspect of the experiment. Additionally, we observed a tendency among participants using LLMs to engage in fewer multi-query search sessions, possibly reflecting uncertainties surrounding LLM capabilities and the perceived affordances associated with LLM interface.

Our findings can extend beyond the geolocation domain, providing initial insights into user interactions with LLMs in real-world applications and prompting more research on human-centered design of LLM interfaces, with a focus on understanding how users form mental models of LLMs. To achieve more useful LLM interfaces, it is necessary to first develop a better understanding of query formulation strategies and behavior. Extensive prior research about traditional search provides a solid foundation for exploring query formulation strategies. Our work presented in this paper begins to extend this research based on the novel capabilities and the conversational nature of LLM-based search; however, more research in this area is needed. The second component of making LLM interfaces more usable is to teach novices how to effectively prompt. Emerging systems like AI Chains~\cite{tongshuang2022aichains}, MemorySandbox~\cite{huang2023memory}, Feedback Buffet~\cite{macneil2023prompt}, and PromptMaker~\cite{10.1145/3491101.3503564} are at the forefront, making LLMs more comprehensible and user-friendly through the use of templates~\cite{10.1145/3491101.3503564, macneil2023prompt} and procedural guidance~\cite{tongshuang2022aichains, huang2023memory,10.1145/3591196.3596818}. These tools are designed to assist novice users in prompt creation by integrating visual problem representation, incorporating partial prompts, and providing user friendly interfaces that facilitate easy iteration based on the LLM output. These advancements represent a leap towards a future in which user interactions with language models become more intuitive, efficient, and user-friendly.

\begin{acks}
Thanks to Andrea Brandt for assisting with the user study. This research was sponsored by the DEVCOM Analysis Center and was accomplished under Cooperative Agreement Number W911NF-22-2-0001. The views and conclusions contained in this document are those of the authors and should not be interpreted as representing the official policies, either expressed or implied, of the Army Research Office or the U.S. Government. The U.S. Government is authorized to reproduce and distribute reprints for Government purposes notwithstanding any copyright notation herein.
\end{acks}


\bibliographystyle{ACM-Reference-Format}
\bibliography{refs}

\end{document}